\documentstyle[epsf,psfig]{elsart}
\begin{document}
\begin{frontmatter}
\title{Transport theory with nonlocal corrections}
\author[Rostock]{K. Morawetz}
\author[Prag]{V\'aclav \v Spi\v cka and Pavel
Lipavsk\'y}
\address[Rostock]{Fachbereich Physik, Universit\"at Rostock,
18051 Rostock, Germany}
\address[Prag]{
Institute of Physics, Academy of Sciences, Cukrovarnick\'a 10,
16200 Praha 6, Czech Republic}

\begin{abstract}
A kinetic equation which combines the quasiparticle drift of Landau's
equation with a dissipation governed by a nonlocal and noninstant
scattering integral in the spirit of Snider's equation for gases is
derived. Consequent balance equations for the density, momentum and
energy include quasiparticle contributions and the second order quantum
virial corrections and are proven to be consistent with conservation laws. 
\end{abstract}
\end{frontmatter}

%\section{Introduction}
The very basic idea of the Boltzmann equation (BE), to balance
the drift of particles with dissipation, is used both in gases, plasmas
and condensed systems like metals or nuclei. In both fields,
the BE allows for a number of improvements which make it possible
to describe phenomena far beyond the range of validity of
the original BE. In these improvements the theory of gases
differs from theory of condensed systems.
In theory of gases, the focus was on so called virial corrections
that take into account a finite volume of molecules, e.g. Enskog included
space non-locality of binary collisions \cite{CC90}. 
In the theory of condensed systems, modifications of the BE are
determined by the quantum mechanical statistics. A headway in
this field is covered by the Landau concept of quasiparticles
\cite{BP91}. There are three major modifications: the Pauli
blocking of scattering channels; underlying quantum mechanical
dynamics of collisions; and quasiparticle renormalization of a
single-particle-like dispersion relation. However, 
the scattering integral of the BE remains local
in space and time. In other words, the Landau theory does not
include a quantum mechanical analogy of virial corrections.
The missing link of two major streams in transport theory
is clearly formulated by Lalo\"e and Mullin \cite{LM90} in their
comments on Snider's equation. Our aim is to fill this gap.
Briefly, here we derive a transport equation that includes
quasiparticle renormalizations in the standard form of Landau's
theory and virial corrections in the form similar to the theory of
gases. ``Particle diameters'' and other non-localities of the scattering
integral are given in form of derivatives of phase shift in binary
collisions \cite{SLMa96,SLMb96}.

%\section{Nonlocal kinetic equation}

A convenient starting point to derive various corrections to the BE
is the quasiparticle transport equation first obtained by Kadanoff
and Baym
\begin{equation}
{\partial f\over\partial t}+
{\partial\varepsilon\over\partial k}
{\partial f\over\partial r}-
{\partial\varepsilon\over\partial r}
{\partial f\over\partial k}
=
z(1-f)\Sigma^<_\varepsilon-zf\Sigma^>_\varepsilon.
\label{tr1}
\end{equation}
Here, quasiparticle distribution $f$, quasiparticle energy
$\varepsilon$ and wave-function renormalization $z$ are functions
of time $t$, coordinate $r$, momentum $k$ and isospin $a$.
The self-energy $\Sigma^{>,<}$ is moreover a function of energy
$\omega$, however it enters the transport equation only by its
value at pole $\omega=\varepsilon$.
The drift terms in the l.h.s of (\ref{tr1}) have the standard form
of the BE except that the single-particle-like energy $\varepsilon$
is renormalized. This is exactly the form of drift visualized
by Landau. The scattering integral in the r.h.s. of (\ref{tr1}) is,
however, more general than expected by Landau, in particular, it
includes virial corrections which emerge for complex self-energies \cite{SL95}.
The self-energy we discuss is constructed from a two-particle T-matrix
in the Bethe-Goldstone approximation (for simplicity, we have left
aside the exchange term)
%\begin{eqnarray}
$\Sigma^<(1,2)
=
T^R(1,\bar 3;\bar 5,\bar 6)
T^A(\bar 7,\bar 8;2,\bar 4)
G^>(\bar 4,\bar 3)
G^<(\bar 5,\bar 7)
G^<(\bar 6,\bar 8),
$
%\label{tr3}
%\end{eqnarray}
which is known to include non-trivial virial corrections \cite{MR95}.
Here, $G$'s are single-particle Green's functions, numbers are
cumulative variables, $1\equiv (t,r,a)$, time, coordinate and isospin.
Bars denote internal variables that are integrated over.
The self-energy as a functional of Green's functions $\Sigma[G]$ is converted into the
scattering integral $\Sigma_\varepsilon[f]$ via the quasiparticle
approximation
%\begin{eqnarray}
$G^>(\omega,k,r,t,a)
=
(1-f(k,r,t,a))2\pi\delta(\omega-\varepsilon(k,r,t,a))$ and $
%\nonumber\\
G^<(\omega,k,r,t,a)
=
f(k,r,t,a)2\pi\delta(\omega-\varepsilon(k,r,t,a))$.
%\label{tr1a}
%\end{eqnarray}
Omitting gradient contributions to collisions one simplifies the
scattering integral, but on cost of virial corrections. Indeed,
the space and time non-locality of the scattering integral is washed
out in absence of gradients. 
To obtain the scattering integral with virial corrections we linearize
all functions in a vicinity of $(r,t)$  using $r^i-r$ and $t^1-t$
as small parameters to second order. 
Then the scattering integral
of equation (\ref{tr1}) results
\begin{eqnarray}
%&&\left(1-f(\upsilon^0_a)\right)
%\Sigma^<_{\varepsilon^0_a}(\upsilon^0_a)-f(\upsilon^0_a)
%\Sigma^>_{\varepsilon^0_a}(\upsilon^0_a)
%\nonumber\\&&
&&\sum_b
\int{dp\over(2\pi)^3}
{dq\over(2\pi)^3}
2\pi\delta
\left(
\varepsilon^0_a+
\varepsilon^3_b-\varepsilon^1_a-\varepsilon^2_b+2\Delta_E
\right)
\nonumber\\
&&\times
|T|^2\!
\left(
\varepsilon^0_a\!+\!\varepsilon^3_b\!-\!\Delta_E,
k\!-{\Delta_K\over 2},\!p\!-\!{\Delta_K \over 2},q,
t\!-\!{1\over 2}\Delta_t,r\!-\!\Delta_r\!
\right)
\nonumber\\
&&\times
\Bigl[
f^1_a
f^2_b
\bigl(1-f^0_a\bigr)
\bigl(1-f^3_b\bigr)-
\bigl(1-f^1_a\bigr)
\bigl(1-f^2_b\bigr)
f^0_a
f^3_b
\Bigr].
\label{tr11}
\end{eqnarray}
Here, $\upsilon^0_a=(k,r,t,a)$,
$\upsilon^1_a=(k-q-\Delta_K,r-\Delta_3,t-\Delta_t,a)$,
$\upsilon^2_b=(p+q-\Delta_K,r-\Delta_4,t-\Delta_t,b)$,
$\upsilon^3_b=(p,r-\Delta_2,t,b)$, and
$\varepsilon^i_a=\varepsilon(\upsilon^i_a)$ and
$f^i_a=f(\upsilon^i_a)$.
One has to keep in mind that form (\ref{tr11}) holds only up to its
linear expansion in $\Delta$'s.
All $\Delta$'s are given by derivatives of the phase shift
\mbox{$\phi={\rm Im\ ln}T^R_{\rm sc}(\Omega,k,p,q,t,r)$},
\begin{equation}
\begin{array}{lclrcl}\Delta_t&=&{\displaystyle
\left.{\partial\phi\over\partial\Omega}
\right|_{\varepsilon_1+\varepsilon_2}}&\ \ \Delta_2&=&
{\displaystyle\left({\partial\phi\over\partial p}-
{\partial\phi\over\partial q}-{\partial\phi\over\partial k}
\right)_{\varepsilon_1+\varepsilon_2}}\\ &&&&&\\ \Delta_E&=&
{\displaystyle\left.-{1\over 2}{\partial\phi\over\partial t}
\right|_{\varepsilon_1+\varepsilon_2}}&\Delta_3&=&
{\displaystyle\left.-{\partial\phi\over\partial k}
\right|_{\varepsilon_1+\varepsilon_2}}\\ &&&&&\\ \Delta_K&=&
{\displaystyle\left.{1\over 2}{\partial\phi\over\partial r}
\right|_{\varepsilon_1+\varepsilon_2}}&\Delta_4&=&
{\displaystyle-\left({\partial\phi\over\partial k}+
{\partial\phi\over\partial q}\right)_{\varepsilon_1+\varepsilon_2}}
\end{array}
\label{8}
\end{equation}
and $\Delta_r={1\over 4}(\Delta_2+\Delta_3+\Delta_4)$.
After derivatives, $\Delta$'s are evaluated at the energy shell
$\Omega\to\varepsilon_1+\varepsilon_2$
The $\Delta$'s are effective shifts and they represent mean values of
various non-localities of the scattering integral. These shifts enter
the scattering integral in form known from theory of gases
\cite{CC90}, however, the set of shifts is larger than the one
intuitively expected. The full set (\ref{8}) is necessary to
guarantee gauge invariance.
One can see that sending all $\Delta$'s to zero, the scattering
integral (\ref{tr11}) simplifies to the one used in the BE for
quasiparticles. The scattering integral is interpreted as
collision at time $t$ and coordinate $r$ in which two particles
(holes) $a$ and $b$ of momenta $k$ and $p$ scatter into final
states of momenta $k-q$ and $p+q$. This interpretation is correct for
the weak-coupling limit $T^R\approx V$, where the phase shift in
dissipative channels vanishes, $\phi=0$, and no virial
corrections appear. With nontrivial $\Delta$'s, the interpretation
has to be slightly modified due to finite collision duration and
finite ``particle diameters''.
For instant potential, the particles $a$ and $b$ enter the collision
at the same time instant (there is no time shift between arguments
$\upsilon^0_a$ and $\upsilon^3_b$) and leave the collision together
(there is no time shift between $\upsilon^1_a$ and $\upsilon^2_b$).
The only time shift $\Delta_t$ is between the beginning and the end of
collision. This time shift is just the collision delay discussed by
Danielewicz and Pratt \cite{DP96}.
Due to the finite duration of the collision, the pair of particles $a$
and $b$ can gain an energy $2\Delta_\omega$ from external fields.
The momentum shift $2\Delta_k$ describes an acceleration the pair of
particles picks up during their correlated motion.

With respect to a general form of the transport equation we have
already fulfilled our task: the quasiparticle transport equation
(\ref{tr1}) with the non-local scattering integral (\ref{tr11})
is our final result. This transport equation has complicated
self-consistent structure: (i) quasiparticle energy depends on
distributions via real part of self-energy, (ii) scattering rate
depends on distributions via Pauli blocking of two-particle
propagation in T-matrix, (iii) $\Delta$'s depend on distributions
also due to Pauli blocking. The same complexity one meets for the
quasiparticle BE, except for neglected $\Delta$'s. In fact, $\Delta$'s
do not represent much of additional work as the T-matrix has to be
evaluated within the BE anyway. 
%\section{Summary}
To summarize, we have derived a Boltzmann-like transport equation
for quasiparticles that includes virial corrections to the scattering
integral via set of shifts in time, space, momentum and energy. We have been able to proof 
conservation laws for density, momentum and energy \cite{SLM96,LSM97}. The presented theory extends the
theory of quantum gases \cite{NTL91,H90} and non-ideal plasma
\cite{BKKS96} to degenerated system.

With respect to numerical implementations the presented theory is as
simple as possible: the scattering integral (\ref{tr11}) includes only
six-dimensional integration as the standard BE, the virial corrections
in form of $\Delta$'s are friendly to simulation Monte Carlo methods.
Numerical tractability of the presented transport equation documents
Ref. \cite{KDB96}, where
space shifts estimated from ground state have been used.

The authors are grateful to P. Danielewicz, D. Kremp and  G. R\"opke
for stimulating discussions. This project was supported by the BMBF
(Germany) under contract Nr. 06R0884, the Max-Planck Society with 
Nr. 44185, the Grant Agency of Czech
Republic under contracts Nos. 202960098 and 202960021,  and the EC Human Capital and
Mobility Programme.

\end{document}